\documentclass[]{spie}  

\usepackage{amsmath,amsfonts,amssymb}
\usepackage{graphicx}
\usepackage[colorlinks=true, allcolors=blue]{hyperref}

\title{Phase-induced amplitude apodization complex mask coronagraph tolerancing and analysis}

\author[a,b]{Justin M. Knight}
\author[a,b,c]{Olivier Guyon}
\author[c]{Julien Lozi}
\author[d]{Nemanja Jovanovic}
\author[b]{Jared R. Males}
\affil[a]{The College of Optical Sciences, University of Arizona, 1630 E. University Blvd., Tucson, AZ 85719, United States}
\affil[b]{Steward Observatory, University of Arizona, 933 N. Cherry Ave., Tucson, AZ 85719, United States}
\affil[c]{Subaru Telescope, National Astronomical Observatory of Japan, National Institutes of Natural Sciences, 650 N. A'ohoku Place, Hilo, HI 96720, United States}
\affil[d]{Caltech Optical Observatories, California Institute of Technology
1200 E. California Blvd., MC 11-17, Pasadena, CA 91125}

\authorinfo{Further author information: (send correspondence to Justin M. Knight)\\E-mail: jknight@optics.arizona.edu}

\begin{document} 
\maketitle

\begin{abstract}
Phase-Induced Amplitude Apodization Complex Mask Coronagraphs (PIAACMC) offer high-contrast performance at a small inner-working angle ($\approx$ 1 $\lambda$/D) with high planet throughput ($>$ 70\%). The complex mask is a multi-zone, phase-shifting mask comprised of tiled hexagons which vary in depth. Complex masks can be difficult to fabricate as there are many micron-scale hexagonal zones ($>$ 500 on average) with continuous depths ranging over a few microns. Ensuring the broadband PIAACMC design performance carries through to fabricated devices requires that these complex masks are manufactured to within well-defined tolerances. We report on a simulated tolerance analysis of a ``toy'' PIAACMC design which characterizes the effect of common microfabrication errors on on-axis contrast performance using a simple Monte Carlo method. Moreover, the tolerance analysis provides crucial information for choosing a fabrication process which yields working devices while potentially reducing process complexity. The common fabrication errors investigated are zone depth discretization, zone depth errors, and edge artifacts between zones.
\end{abstract}


\section{INTRODUCTION}
\label{sec:intro}  

Phase-Induced Amplitude Apodization Complex Mask Coronagraphs (PIAACMC) offer high-contrast performance at a small inner-working angle (IWA $\approx$ 1 $\lambda$/D) for direct imaging of exoplanets while maintaining high planet throughput ($>$ 70$\%$)\cite{CMC}. The PIAACMC architecture works well for both ground- and space-based telescopes; it reaches required contrast levels for imaging earth-like planets ($\approx$ 10$^{-10}$) because it is robust against complicated telescope pupil architectures which include segment gaps, spiders, and a secondary obscuration. The complex amplitude focal plane mask (FPM), or complex mask, is a multi-zone, phase-shifting mask. This device admits several improvements over the previous PIAA coronagraph concept: easier PIAA optics fabrication thanks to milder aspheric optics profiles and broadband performance of up to a 20\% wavelength bandwidth. Recent complex mask designs are described by tiled hexagons of equal size which vary in depth over a fraction of the point spread function (PSF) core (typically a 2-3 $\lambda$/D radius); the combined effect of these zones is to induce optical path length delays which destructively interfere light at multiple wavelengths simultaneously in the science PSF, resulting in a deep achromatic null. As PIAACMC designs have been pushed to higher performance, complex masks have become increasingly complicated, containing many ($>$ 500) zones over a continuous range of depths covering just a few microns. Because of this, complex masks can be challenging to fabricate. Knowing how well the masks can be made versus how well they have to be made is important; understanding this difference allows us to fabricate devices which will perform well without over-complicating the engineering process to make them. To this end, we perturb the complex mask of a simulated PIAACMC design and measure the resulting contrast sensitivity to each input parameter. Describing input parameters of interest is easily understood once a fabrication path is chosen.

\section{COMPLEX MASK FABRICATION}
\label{sec:fab}
\begin{figure} [ht]
   \begin{center}
   \begin{tabular}{c} 
   \includegraphics[height=7cm]{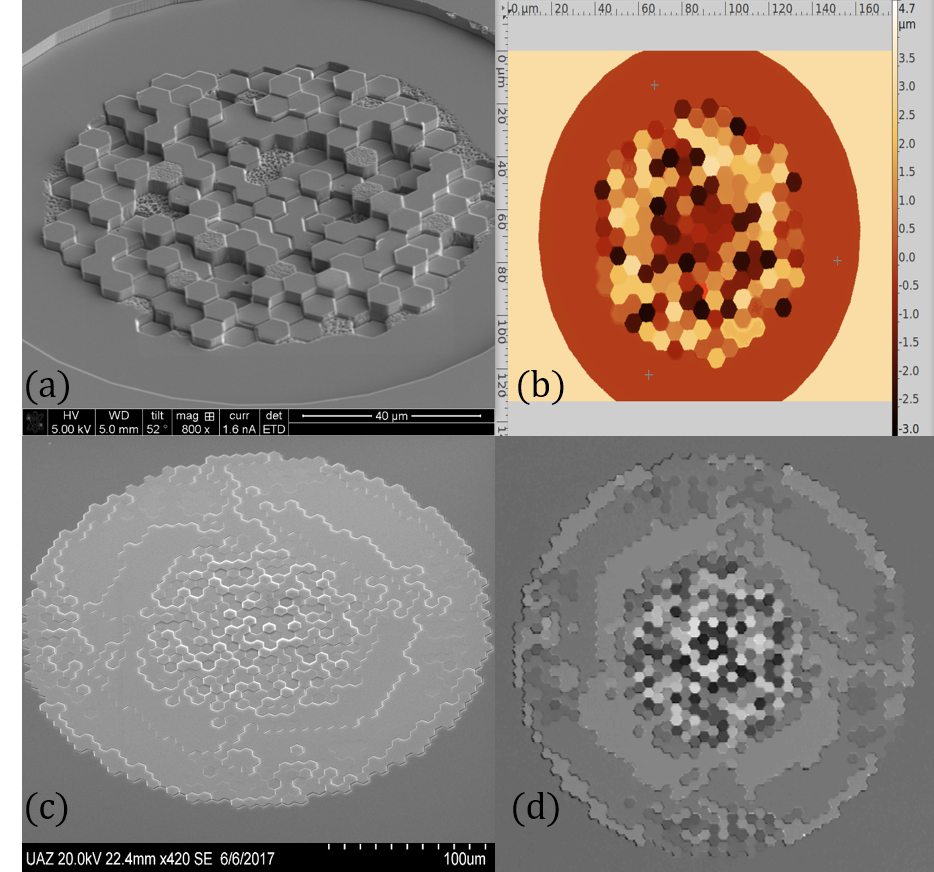}
	\end{tabular}
	\end{center}
   \caption[example] 
   { \label{fig:FABMASKS}
   (a) A SiO2 complex mask fabricated by Christopher Alpha from Cornell NanoScale Science and Technology Facility (CNF). (b) An interferometer image capturing the zone heights present in the mask in (a). (c) A Si complex mask fabricated at Arizona State University in conjunction with the University of Arizona. (d) An interferometric image of the mask in (c).
}
   \end{figure}

Examples of successfully fabricated focal plane masks at high-contrast imaging testbeds include those made and tested for the WFIRST mission at the High-Contrast Imaging Testbed at NASA JPL\cite{WFIRST_PIAA}, the Ames Coronagraph Experiment at NASA Ames\cite{FPM_PIAA}, and the Subaru Coronagraphic Extreme Adaptive Optics (SCExAO) instrument at the Subaru telescope\cite{UofAFAB,SCExAO}. In each case, different fabrication processes were employed, ranging from grayscale direct-writing with an e-beam lithography tool, to more standard microfabrication processes like binary optics. Binary optics\cite{Lithography} is the process by which most semiconductor devices are commonly made: the micron scale device of interest, typically containing structures of varied depths, can be approximated by a series of equally spaced depths. These depths are achieved by patterning a photosensitive chemical, or photoresist, onto a material substrate which will hold the pattern indefinitely. The patterns are transferred by using binary masks (commonly made of chrome on glass) to expose or block light. The photoresist is then developed, leaving some substrate material exposed. Chemical etching can then be used to create the first depth. Each subsequent photolithographic and etch step proceeds at half of the previous depth. After each step in the process, a total of 2$^{n}$ depths approximate the device depths in the substrate. Fig. \ref{fig:FABMASKS} depicts the scanning electron microscope and white-light interferometric images of the complex masks fabricated using binary optics for the SCExAO instrument.

These masks have been delivered to or installed in the instrument. While these masks may not directly benefit from a tolerancing analysis, they offer a means to compare simulated and measured data against testbed contrast performance. Moreover, with their recent installation, the masks can be tested on-sky to determine how well a first round of complex masks perform on a ground-based telescope. In addition, other instruments and testbeds gearing up for high-contrast imaging such as MagAO-X\cite{MAGAO-X} and SPEED\cite{SPEED} will benefit from the lessons learned examining the PIAACMCs at SCExAO.

\subsection{Fabrication errors}
There are several types of fabrication errors associated with binary optics. We will consider a few which are very common, that is, difficult to avoid during a fabrication run. First, changing a continuous surface (or number of depths) to a discrete number of levels is known as the discretization error. The next error we consider is etching to the wrong zone depth. Finally, lateral mis-registrations between binary mask exposures creates overlap between bordered zones, causing edge artifacts. 

\section{PIAACMC SIMULATION}

PIAACMC design and optimization is performed using the software package COFFEE: Coronagraph Optimization For Fast Exoplanet Exploration\cite{COFFEE}. COFFEE is written in C and is organized into modules to perform optical system propagations and optimization routines for PIAA-style coronagraphs. Briefly, the software uses Fresnel diffraction calculations and discrete Fourier transforms to propagate between elements of a PIAACMC in an unfolded optical configuration. COFFEE constructs PIAA optics surfaces, Lyot stop shapes, and complex mask solutions to produce a high-performance, broadband coronagraph capable of achieving aggressive IWAs ($\approx$ 1 $\lambda$/D) while maintaining both high throughput and contrast. The software can accept additional wavefront error modes to reduce sensitivity to well-known problems such as tip-tilt due to telescope pointing stability. While all of the routines are present to handle a complex mask tolerance analysis, or even a full element-wise tolerance analysis which considers the effects of fabrication, the code base has not been updated to perform this type of analysis by the original author. Moreover, contributors outside the original author face a steep learning curve without prior coding experience in C. The combination of these reasons pushes us to use a high-level language with built-in routines that can facilitate the development of a tolerancing analysis method which considers performance sensitivity to the fabrication processes used to create a complex mask.

\subsection{From COFFEE to MATLAB: perturbing complex masks}
The purpose of the work in MATLAB is to simulate microfabrication process errors present in a fabricated complex mask and analyze their effects on on-axis contrast performance. To begin, we assume an input wavefront from COFFEE which describes the pupil after light has passed through the PIAA optics at each wavelength sample across the design bandwidth. Next, we employ a matrix Fourier transform to the input wavefront which samples the PSF core across the complex mask in greater detail; this allows us to more carefully examine fabrication errors. The complex mask solution is then transformed to match the fabrication process described in Sec. \ref{sec:fab}. This process begins by converting the continuous range of zone depths into a discrete number of equally spaced depths. This information is used to generate $n$ binary exposure patterns, or layers, which are used to achieve the 2$^{n}$ depths now representing the zone heights. The layers are 2D maps which mask off zones which are not specified by the corresponding depth. Each of these layers are then scaled, shifted, and added together to create an approximate version of the complex mask with some common fabrication errors present. For example, scaling a layer will produce an etch depth error across each zone to be exposed, while shifting a layer mask by a few pixels is equivalent to lateral uncertainty when aligning one layer mask with another in successive lithographic exposures. Note that an additional layer is required to create both positive and negative zone depths, known as the bipolar layer. For example, this is the recessed circle surrounding the mask in Fig. \ref{fig:FABMASKS}(a). This step can add error during the actual fabrication of a complex mask device, but for simplicity we assume the error in depth and layer shifts from the other layers has absorbed any ill effects born from the bipolar layer.

Following the creation of the fabricated complex mask, it is applied to the up-sampled PSF core; to return to the pupil plane, we use Babinet's principle in conjunction with another matrix Fourier transform. From there, the effect of the complex mask has been computed and is subtracted from the input pupil wavefront. The angular spectrum method is applied to this post complex mask pupil to propagate to each Lyot stop location, where the corresponding Lyot mask is applied. Finally, a Fourier transform is used to compute the coronagraphic PSF.

\subsubsection{Sensitivity metrics}
The output of the MATLAB simulation is a coronagraphic PSF which is used to compute contrast curves and average contrast over a region of interest. We define the contrast curve metric as the radially-averaged flux ratio between the coronagraphic and non-coronagraphic PSFs averaged over wavelength. Additionally, we compute the average contrast over a C-shape between two and six $\lambda$/D in a similar manner. While other metrics are interesting to calculate in this scheme, we expect contrast to be a good indicator of how well a complex mask must be fabricated to achieve design-level performance because the calculation uses information from the PSF core and surrounding structure where the complex mask directly affects light.

\subsection{``Toy" LUVOIR model}
\begin{figure} [ht]
   \begin{center}
   \begin{tabular}{c} 
   \includegraphics[height=7cm]{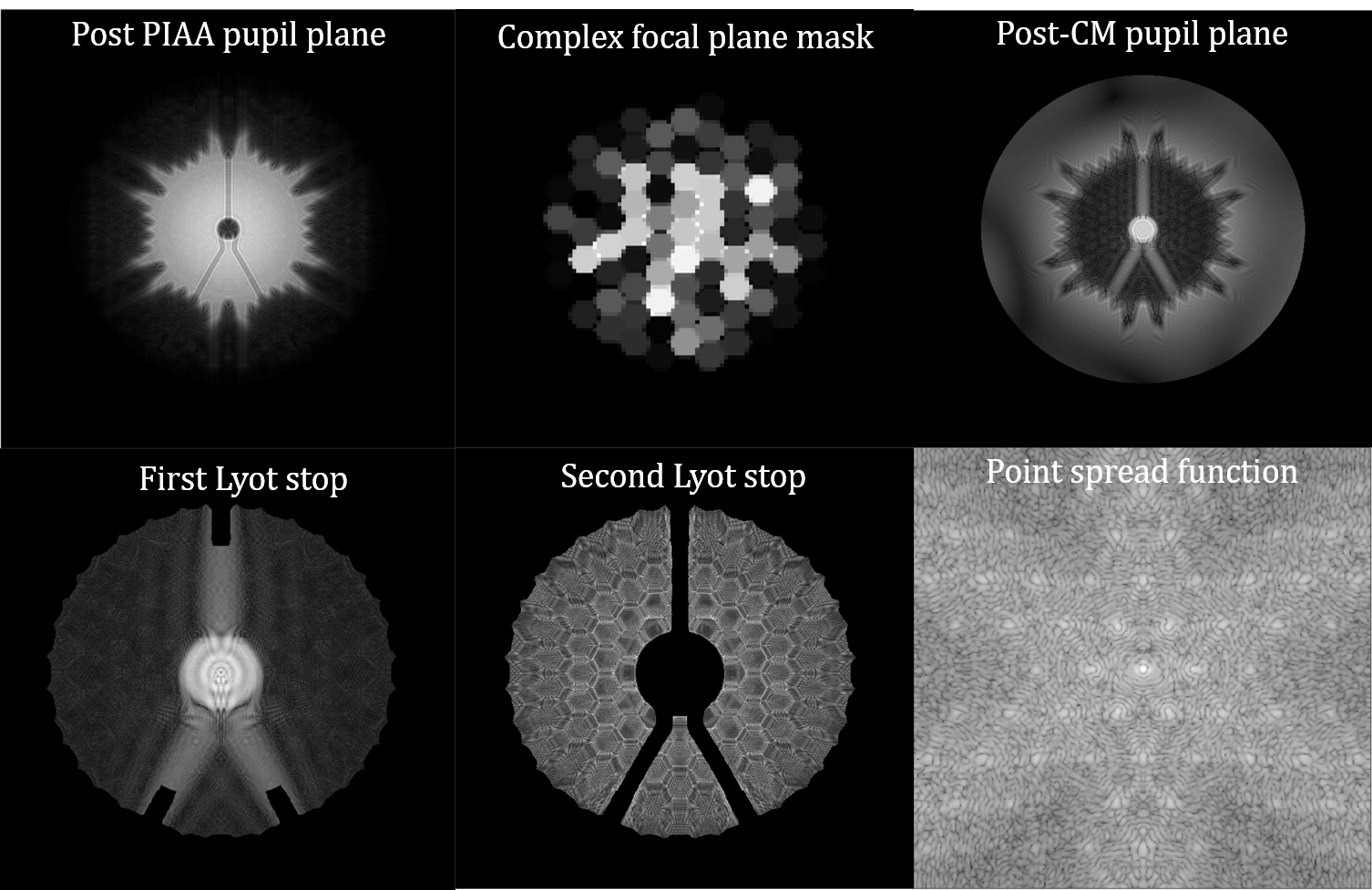}
	\end{tabular}
	\end{center}
   \caption[example] 
   { \label{fig:Toy}
   Each field or element image is scaled, at one wavelength, to show more features than on a linear scale. Top row, starting from the left: pupil plane post PIAA optics; phase of the complex mask solution resulting in the best broadband performance; conjugate pupil plane after light has gone through the complex mask. Bottom row, starting from the left: application of the first Lyot stop; application of the second Lyot stop; the coronagraphic PSF.
}
   \end{figure}

To demonstrate the tolerancing process, we use a ``toy" PIAACMC design. This model is a very preliminary design case for a LUVOIR-type pupil architecture\cite{LUVOIR} to be designed, fabricated and tested at the ACE testbed. The complex mask contains 85 distinct zones over a four $\lambda$/D diameter at a central wavelength of 1300nm. The post-PIAA pupil plane wavefront from the COFFEE optimization process is shown at the top-left of Fig. \ref{fig:Toy}. The image to the right of the input wavefront is the best solution for the complex mask depicted at the sampling used to compute zone depths during the optimization process. Each wavefront following the complex mask in order from left to right follows the process described above: the pupil after the complex mask is applied, application of the first Lyot stop, application of the second Lyot stop, and the resulting PSF. Note the Lyot stops do not reside in conjugate pupil plane locations; the optimization process searches a range around the conjugate Lyot stop plane where multiple Lyot stops can be used to best mask diffractive features due to spiders, segment gaps, and the secondary obscuration.

\begin{figure} [ht]
   \begin{center}
   \begin{tabular}{c} 
   \includegraphics[height=7cm]{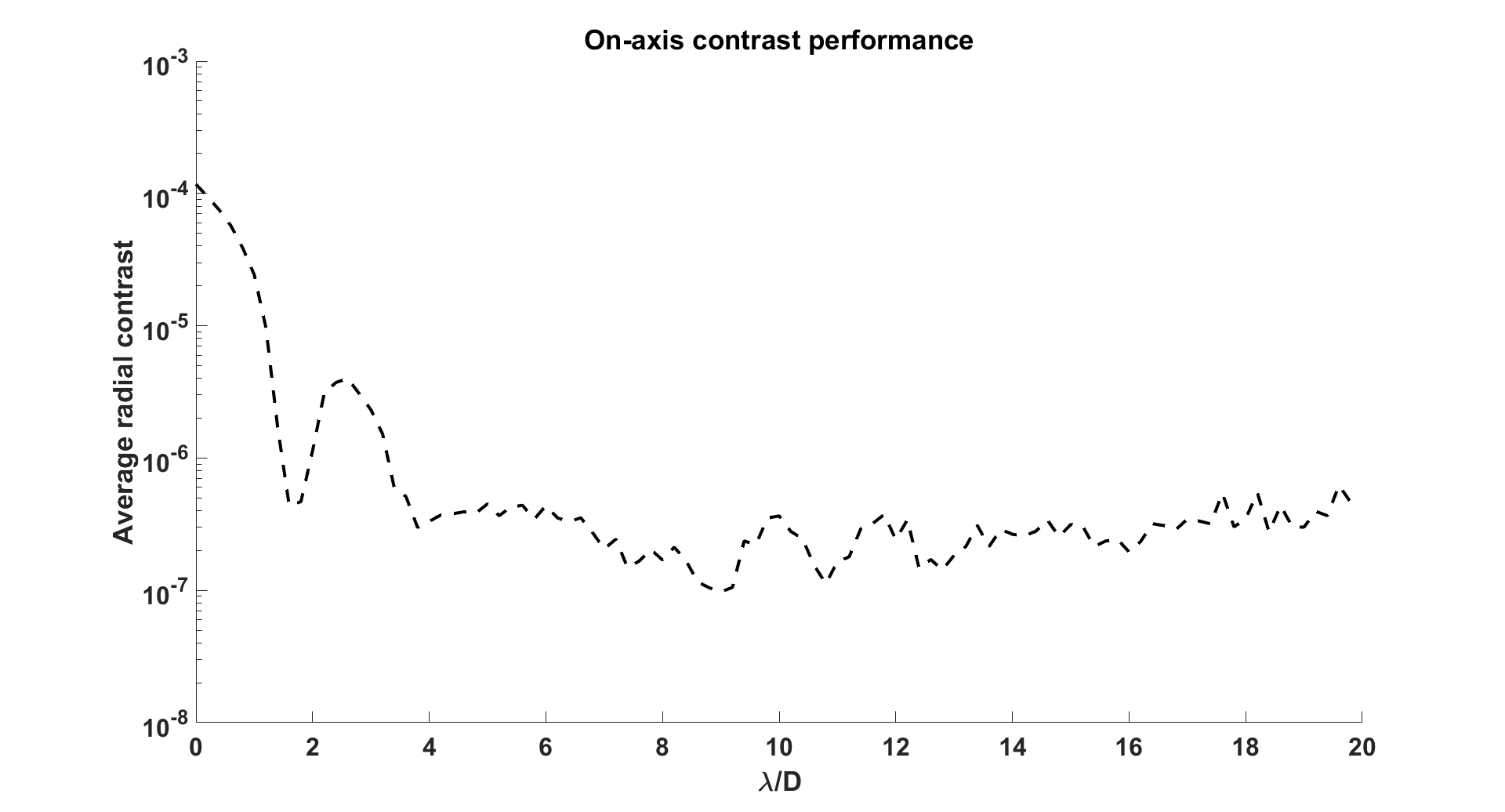}
	\end{tabular}
	\end{center}
   \caption[example] 
   { \label{fig:Design_contrast}
   The contrast curve is plotted on a logarithmic scale. The estimated IWA is about 1.7 $\lambda$/D, resulting in a contrast of approximately 10$^{-6}$. On-axis contrast performance is computed by considering a tilted point source which does not go through the complex mask. The resulting PSF is treated as the denominator of the flux ratio.
}
   \end{figure}

The on-axis contrast curve resulting from this design is shown in Fig. \ref{fig:Design_contrast}. With an average contrast of $\approx$ 10$^{-6}$, this is not an acceptable PIAACMC for a space telescope, but it does not need to be for our purposes. It is enough to use this design to demonstrate that our methods of perturbing the complex mask and propagating the effects through to the PSF yield meaningful results to define tolerance specifications for a complex mask given available fabrication processes.

\section{SIMULATION RESULTS}
We present two simulations: the first isolates the effect of the discretization process on the contrast curves, while the second assumes a level of discretization and introduces shifts and depth errors in each layer. The latter simulation draws numbers from a uniform distribution which reflect - to some extent - common variables encountered during a fabrication run. The layer shifts, which occur with respect to the first layer, are drawn as random integers which shift by whole pixel amounts in the x- and y- directions. Because there are only 85 hexagons across the PSF in this design, they are quite easy to fabricate within the alignment limitations of standard lithographic exposure tools. Each pixel shift magnitude is chosen at random so that the bordering hexagonal zones have similar proportional area of overlap as in the case of an actual fabrication run with hundreds of zones as discussed in Sec. \ref{sec:fab}. Meantime, the depth errors are given similar limitations which result from likely outcomes of a fabrication run. As mentioned in the standard scheme for etching, each subsequent etch depth is half as much as the previous. As the depths approach tens of nanometers or less, as is the case for the mask in Fig. \ref{fig:FABMASKS}(c), it becomes difficult to both control the etch rate and accurately measure how much material was etched with metrology tools. To reflect this, we increase the percent each layer depth can be off by 5\% for each layer.

\begin{figure} [ht]
   \begin{center}
   \begin{tabular}{c} 
   \includegraphics[height=7cm]{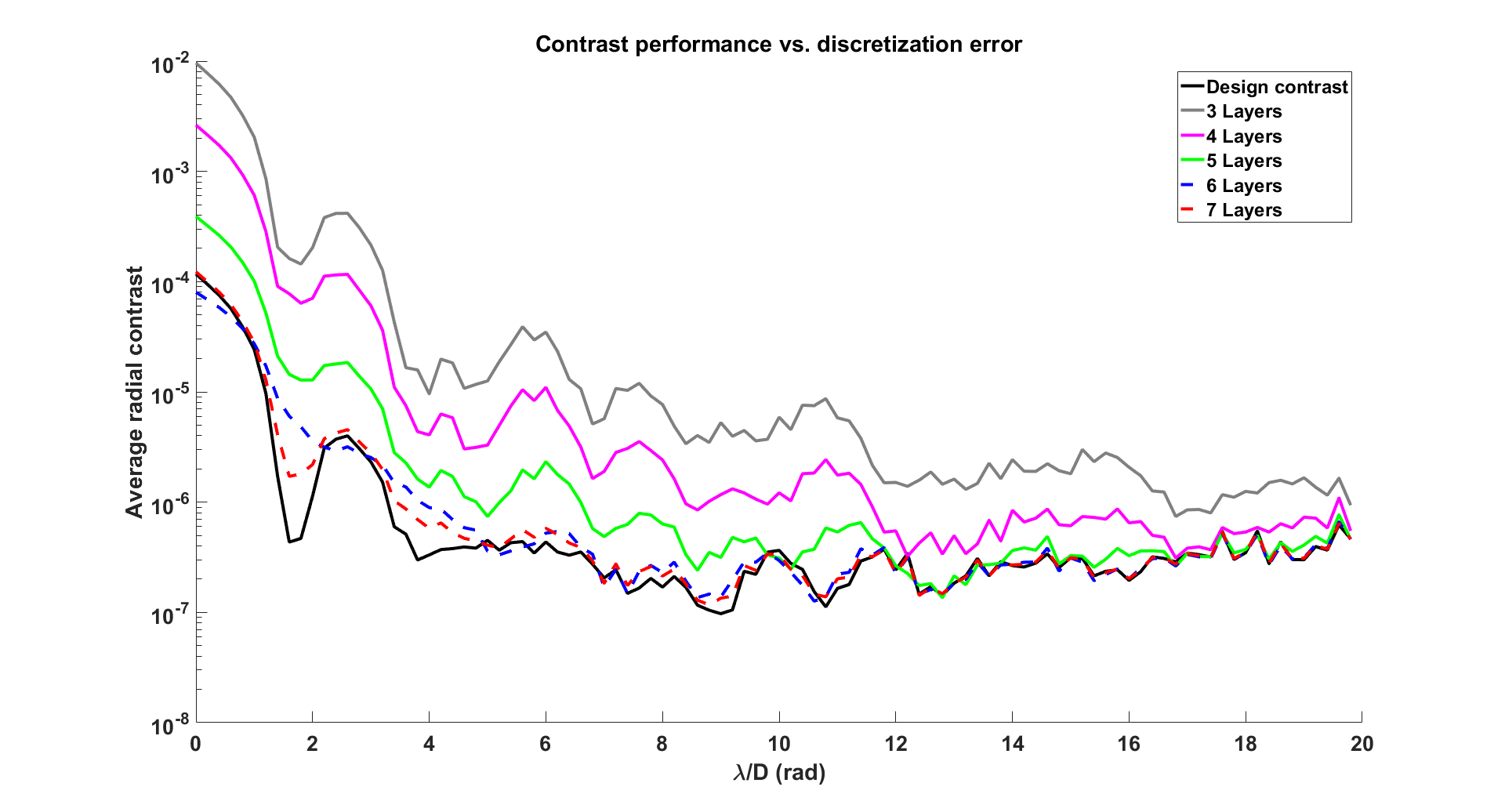}
	\end{tabular}
	\end{center}
   \caption[example] 
   { \label{fig:discrete_contrast}
   The on-axis contrast performance is calculated for each mask. Each curve shifting up along the y-axis represents a worse approximation to the designed number of depths. Two points in each of the computed curves around the first minimum are linearly interpolated, resulting in an artificial mismatch to the design curve. This occurs because no points in the PSF are sampled during the contrast curve generation. Moreover the six and seven layer curves do not completely agree with the design curve because of slight differences in complex mask zone map generation in moving from COFFEE to MATLAB.
}
   \end{figure}

The results of the first simulation are shown in Fig. \ref{fig:discrete_contrast}. The black curve is the design contrast, while each of the other curves corresponds to the increasingly crude approximations of the discretization process when using binary optics. It is clear that each time a coarser approximation is taken, i.e. a layer is reduced, the contrast curve shifts up along the y-axis. This is an expected behavior because as the number of layers is reduced, more complex mask zones become degenerate with one another; the net effect being the loss of chromatic nulling as the PSF structure becomes less modulated at each wavelength. Another way of thinking about this is that as the number of zones goes to one, the complex mask becomes a single phase-shifting mask, largely capable of working at only the central wavelength.

\begin{figure} [ht]
   \begin{center}
   \begin{tabular}{c} 
   \includegraphics[height=7cm]{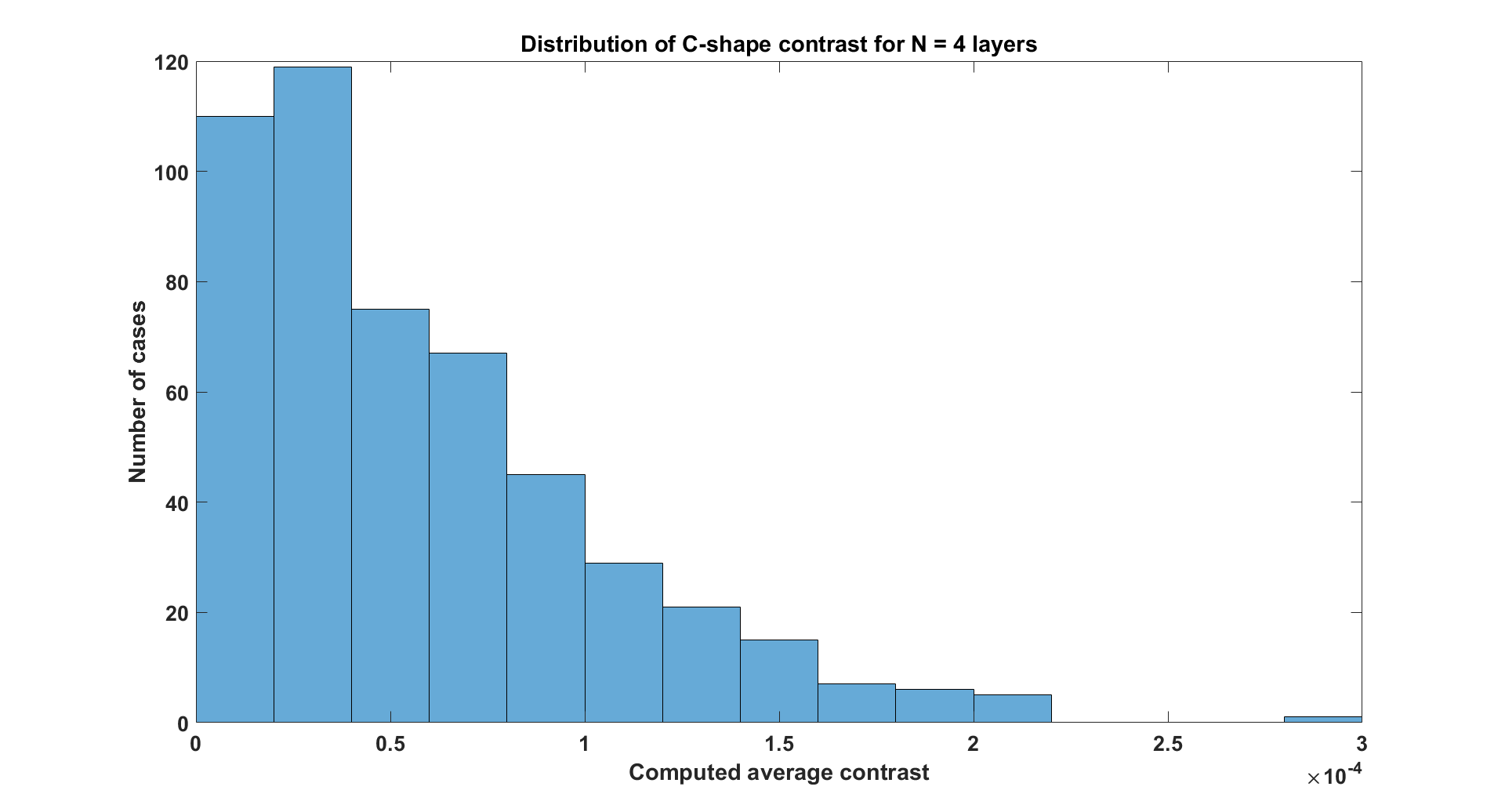}
	\end{tabular}
	\end{center}
   \caption[example] 
   { \label{fig:4Layer_Monte}
   A Monte Carlo simulation of a complex mask consisting of 2$^{4}$ = 16 discrete depths and 500 random realizations of layer shifts and depth errors. The histogram follows an exponential distribution, indicating that most masks are not significantly degraded by these types of errors.
}
   \end{figure}

\begin{figure} [ht]
   \begin{center}
   \begin{tabular}{c} 
   \includegraphics[height=7cm]{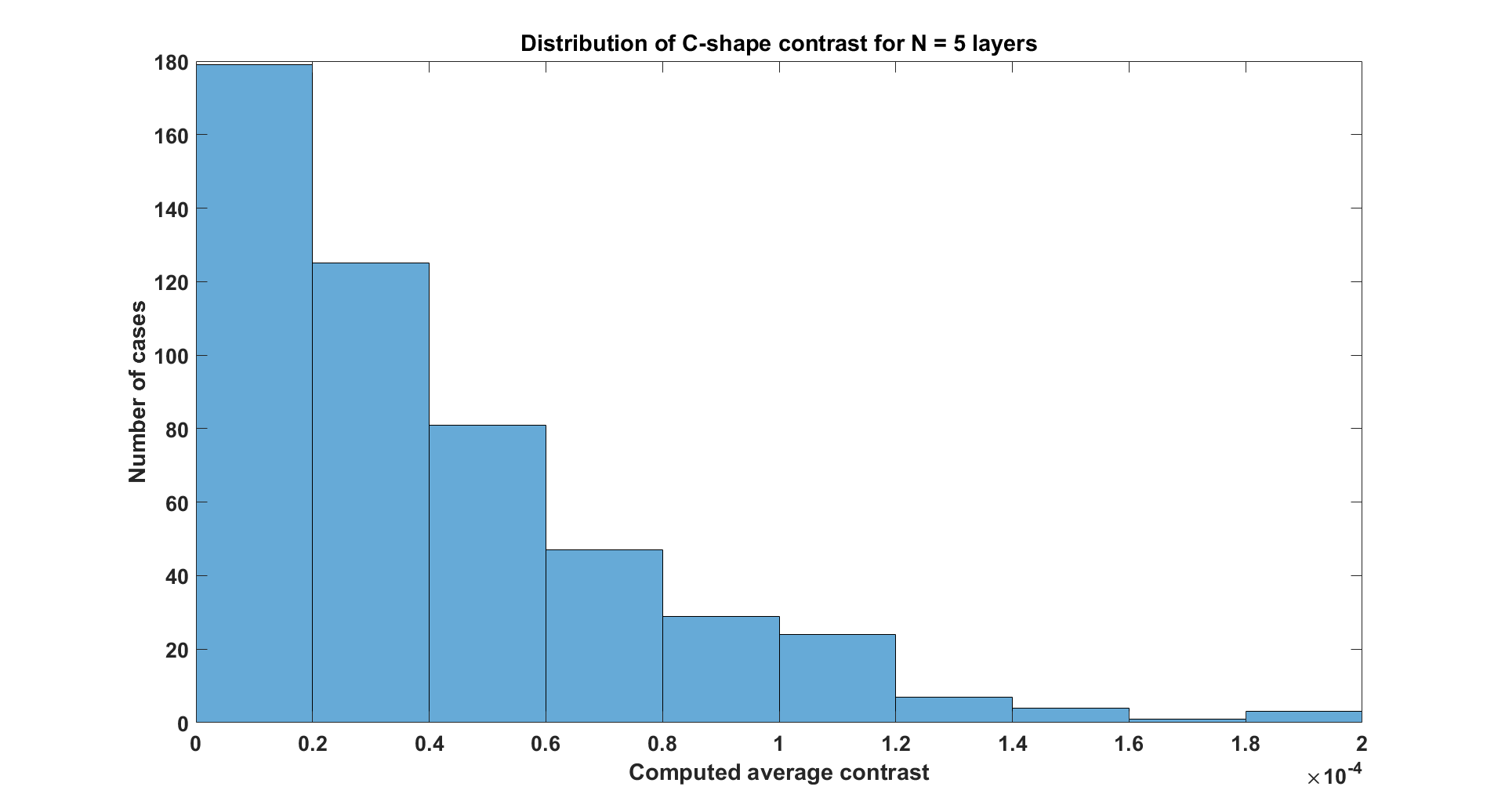}
	\end{tabular}
	\end{center}
   \caption[example] 
   { \label{fig:5Layer_Monte}
   A Monte Carlo simulation of a complex mask consisting of 2$^{5}$ = 32 discrete depths and 500 random realizations of layer shifts and depth errors. The histogram follows an exponential distribution, indicating that many of the 500 cases retain performance to within a factor of 10.
}
   \end{figure}

The second simulation details the effects of some of the limitations which occur during each lithographic and etch step of the fabrication process assuming a number of exposures. The cases shown in Fig. \ref{fig:4Layer_Monte} and Fig. \ref{fig:5Layer_Monte} correspond to four and five exposures respectively. Each distribution is computed from 500 uniformly distributed random samples, the result in each case being an exponential distribution of the average contrast. The calculated average contrast for four layers is \(2.59\cdot10^{-5}\), while for five layers it is \(4.66\cdot10^{-6}\). The average of the average contrasts computed from the four layer Monte Carlo simulation is \(5.81\cdot10^{-5}\) while the five layer simulation sits at \(4.03\cdot10^{-5}\). These numbers suggest that all else being equal, if the fabrication effort is performed carefully and considerately during each process step, i.e. there is a mature, repeatable process to align each layer to within lateral tolerances of the lithographic tool as well as control each etch step repeatably, that on average, the resulting fabricated complex mask will be worse than designed, yet still capable of decades of suppression.
   
These preliminary Monte Carlo simulations are the first steps to indicating the performance limitations imposed by the fabrication process of binary optics. We find that the most detrimental source of performance degradation comes directly from the fabrication process itself. Further quantitative analysis is necessary to determine any other disparaging factors on contrast performance such as zone shape and size bias, rotational alignment between layers, and alignment of the complex mask itself over the PSF core. Moreover, setting tolerances to ensure a desired level of contrast performance is met is an essential next step.

\section{FUTURE WORK}
\label{sec:future}

The findings of the ``toy" simulation indicate a major sensitivity of contrast to binary optics, but we can go much further by using this tool to characterize the performance of already fabricated masks, particularly those present in the SCExAO testbed. Moving forward, we will model the performance of fabricated masks in Fig. \ref{fig:FABMASKS} using measured data in addition to characterizing them in the testbed and on-sky. This will provide invaluable feedback to the next generation of fabricated complex masks for PIAACMC architectures.

\acknowledgments 

This research is supported in part by a NASA TDEM grant and NSF MRI Award \#1625441 (MagAO-X). A portion of the fabrication work was done in the ASU Nanfab Facility. A portion of the fabrication work was done in the Cornell NanoScale Science and Technology Facility.

\bibliography{report} 

\begin{thebibliography}{10}

\bibitem{CMC}
Guyon, O., Martinache, F., Belikov, R., and Soummer, R., ``High performance
  piaa coronagraphy with complex amplitude focal plane masks,'' {\em The
  Astrophysical Journal Supplement Series}~{\bf 190}(2),  220 (2010).

\bibitem{WFIRST_PIAA}
Kern, B., Guyon, O., Belikov, R., Wilson, D., Muller, R., Sidick, E.,
  Balasubramanian, B., Krist, J., Poberezhskiy, I., and Tang, H.,
  ``Phase-induced amplitude apodization complex mask coronagraph mask
  fabrication, characterization, and modeling for wfirst-afta,'' {\em Journal
  of Astronomical Telescopes, Instruments, and Systems}~{\bf 2},  2 -- 2 -- 8
  (2016).

\bibitem{FPM_PIAA}
Newman, K., Conway, J., Belikov, R., and Guyon, O., ``Focal plane phase masks
  for piaa: Design and manufacturing,'' {\em Publications of the Astronomical
  Society of the Pacific}~{\bf 128}(963),  055003 (2016).

\bibitem{UofAFAB}
Knight, J.~M., Brewer, J., Hamilton, R., Ward, K., Milster, T.~D., and Guyon,
  O., ``Design, fabrication, and testing of stellar coronagraphs for exoplanet
  imaging,'' {\em Proc.SPIE}~{\bf 10400},  10400 -- 10400 -- 11 (2017).

\bibitem{SCExAO}
Lozi, J., Guyon, O., Jovanovic, N., Pathak, P., Skaf, N., Sahoo, A.,
  Martinache, F., Singh, G., Kuhn, J.~G., Serabyn, E., Murakami, N., Nishikawa,
  J., Snik, F., Doelman, D.~S., Mazin, B., Walter, A., Kudo, T., Groff, T.~D.,
  Chilcote, J., Kasdin, J., Tamura, M., and Currie, T., ``Scexao: new
  high-performance coronagraphs ready for science,'' {\em Proc. SPIE}
  (10706-207) (2018).
\newblock Publication pending.

\bibitem{Lithography}
Mack, C.,  [{\em Introduction to Semiconductor
  Lithography}{\nolinebreak\hspace{0.1em}]}, John Wiley and Sons, Ltd (2007).

\bibitem{MAGAO-X}
Males, J.~R., Close, L.~M., Miller, K.~L., Schatz, L., Lumbres, J., Doelman,
  D.~S., Snik, F., Guyon, O., Knight, J., Rodack, A.~T., Morzinski, K.~M.,
  Jovanovic, N., Lozi, J., Mazin, B.~A., Ireland, M.~J., Kenworthy, M.~A.,
  Keller, C.~U., Gorkom, K.~V., Long, J.~D., Hedglen, A.~D., Kautz, M.~Y., and
  Bohlman, C., ``Magao-x: project status and first laboratory results,'' {\em
  Proc. SPIE} (10703-9) (2018).
\newblock Publication pending.

\bibitem{SPEED}
Martinez, P., Beaulieu, M., Guyon, O., Gouvret, C., Dejonghe, J., Preis, O.,
  and Abe, L., ``Design, specification and manufacturing of a piaacmc for the
  speed testbed,'' {\em Proc. SPIE} (10702-148) (2018).
\newblock Publication pending.

\bibitem{COFFEE}
Guyon, O., ``Coffee: Coronagraph optimization for fast exoplanet exploration.''
\newblock https://github.com/coffee-org/coffee.

\bibitem{LUVOIR}
Bolcar, M.~R., Aloezos, S., Bly, V.~T., Collins, C., Crooke, J., Dressing,
  C.~D., Fantano, L., Feinberg, L.~D., France, K., Gochar, G., Gong, Q., Hylan,
  J.~E., Jones, A., Linares, I., Postman, M., Pueyo, L., Roberge, A., Sacks,
  L., Tompkins, S., and West, G., ``The large uv/optical/infrared surveyor
  (luvoir): Decadal mission concept design update,'' {\em Proc.SPIE}~{\bf
  10398},  10398 -- 10398 -- 24 (2017).

\end{thebibliography}
\bibliographystyle{spiebib} 

\end{document}